# Heteroepitaxy of FCC-on-FCC Systems of Large Misfit

Paul Wynblatt[1*], Dominique Chatain[2], Ulrich Dahmen[3]

[1]Department of Materials Science and Engineering, Carnegie Mellon University, Pittsburgh, PA 15213, USA

[2]Aix-Marseille Univ, CNRS, CINaM, 13009 Marseille, France

[3]NCEM-Molecular Foundry, LBNL, Berkeley, CA 94270, USA

* corresponding author

## Abstract

To understand the effects of lattice mismatch on heteroepitaxial growth, we have studied the equilibrium structure and orientation relationships (ORs) of FCC films grown epitaxially on FCC substrates, using molecular dynamics simulations in conjunction with embedded atom method potentials. Three film/substrate systems have been investigated, namely: Ag on Cu, Ag on Ni and Pb on Al. These systems cover a significant range of lattice mismatch, from 12.6% for Ag/Cu to 21.8% for Pb/Al. For each system, the ORs of films on six different substrate orientations, namely: (100), (511), (311), (211), (322) and (111), have been investigated. Films on these susbstrates cover a gradual transition from the oct-cube orientation relationship, which occurs only on (100) substrates, to the heterotwin orientation relationship, which often occurs on (111) substrates. It is found that the resulting ORs vary systematically with substrate orientation, but that the pattern of variation is almost identical for all three systems, and therefore largely independent of mismatch. However, the manner in which mismatch is accommodated does depend on the magnitude of mismatch. Simulations point to an important role for edge-to-edge matching and defects such as stacking faults. An analysis of these results in terms of transformation strains highlights the distinction between the ORs, which are largely independent of mismatch, and the local interfacial structure, which changes directly with mismatch.

# 1. Introduction

Heteroepitaxial growth of thin films is of great scientific and technological importance. The precise control of multilayer films in lattice-matched heteroepitaxial growth is essential for semiconductor devices ranging from computers to solid-state lighting [1]. However, for systems with large mismatch, our understanding of heteroepitaxy remains inadequate. Even the simple case of FCC films on FCC substrates remains poorly understood. When the mismatch is large, the orientation relationship (OR) is often not the simple cube-on-cube alignment. Recent experimental work on the Ag/Ni system, with a ~16% mismatch, discovered a systematic pattern of orientation relationships for Ag films grown on two hundred different crystallographic planes of Ni substrates [2]. It was found that on roughly half of all possible Ni surface orientations, Ag films grew in cube-on-cube OR, but for the other half, the OR varied continuously with substrate orientation. Molecular dynamics (MD) simulations on a subset of 12 different Ni substrates replicated this behavior well [3,4], and a crystallographic analysis implied that the interface contained an eigenvector of the transformation matrix of Ni to Ag [3]. While that work provided valuable insights on heteroepitaxy in a simple large-mismatch FCC/FCC system, it raised new issues that have yet to be addressed. Chief among these is the question of how mismatch affects the OR and the interface structure. The present paper addresses these questions using MD simulations.

To investigate the role of lattice mismatch on the heteroepitaxy between A-films on B-substrates of different orientations, we compare the ORs and interface structures of three FCC/FCC alloy systems: Ag/Cu, Ag/Ni and Pb/Al. These three binary systems are characterized by a low mutual solubility and lattice mismatch, $\rho$, of $a_{Ag}/a_{Cu}=1.126$, $a_{Ag}/a_{Ni}=1.162$ and $a_{Pb}/a_{Al}=1.218$ (where $a_i$ is the lattice constant of component i). The choice of these systems minimizes chemical effects due to mutual solubilities while covering a large range of mismatch in metals that all share the same FCC crystal structure.

MD calculations allow a determination of the OR of a crystalline film on a given substrate orientation without the need to formulate any *a priori* assumptions about the equilibrium OR of the film. By starting with an initial "amorphous" film in which the atoms are randomly distributed, we avoid any bias in the equilibrated OR. This is preferable to the approach of assuming specified abutting interfacial planes, which is usually employed to predict ORs (e.g. see [5,6]). Unlike MD, that approach is not able to explore the whole range of possible configurations, and may miss the lowest energy OR.

Experimental studies on the other hand have generally been limited in scope to low-index or vicinal substrate orientations [7]. Systematic studies of ORs on large sets of substrate surface orientations by combinatorial substrate epitaxy, such as our experimental investigation of Ag films on Ni substrates [2], have previously only been performed on oxide-on-oxide systems (see e.g. [8,9]).

**2. Summary of the experimental and calculated ORs of Ag on Ni(hkl)**

With the advent of electron backscatter diffraction (EBSD) in a scanning electron microscope, it has become possible to determine the relative orientations exhibited by a particular film material, A, on a given *polycrystalline* substrate, B, thereby enabling the study of the ORs that develop on a wide range of substrate orientations, under identical experimental conditions (see, for example, refs [2,8,9]). In two recent studies [2,4], we have compared experimental determinations of the ORs of Ag films/islands[*] equilibrated on some 200 different Ni substrate orientations [2], with ORs obtained by computer simulation of 12 distinct Ni substrate orientations [4]. The simulations were performed by MD in conjunction with embedded atom method (EAM) potentials [10] and showed excellent agreement with the experimental results [2].

Here we provide a summary of the experimental results, since the 200 different substrate orientations examined experimentally yield a far more complete picture of the ORs displayed by Ag films on Ni substrates than could be obtained from the smaller sample of substrate orientations for which simulations were performed. The results are presented schematically in Fig. 1, in the form of partial stereographic projections, one for Ni and one for Ag. We use the stereographic triangle with corners at (100), (111) and (110) orientations as the FCC Standard Stereographic Triangle (SST) for the surface orientations of Ni substrates. The orientations (hkl) in this triangle are characterized by all-positive Miller indices, with h≥k≥l≥0. The interfacial orientations (HKL) of the Ag films are displayed on a similar SST, as well as an appended triangle with corners at (210), (11$\bar{1}$) and (110) (colored pink in Fig. 1).

In our previous experimental study [2], the Ni substrate surface orientations were distributed essentially uniformly in the SST, but the Ag ORs differed depending on the Ni substrate orientation. In the (111)-(110)-(210) region of Ni substrate orientation space (shown in dark blue in the SST of Ni of Fig. 1) the Ag films adopted a cube-on-cube OR (*OR C*). A singular

---

[*] During low-temperature deposition, the film initially nucleates as epitaxial islands. Continued deposition leads to coalescence of these islands to form single-crystal films covering each substrate grain. During subsequent high-temperature annealing, the epitaxial films break up into arrays of similarly oriented islands by solid state dewetting. These islands are then free to equilibrate, subject to the same interface/surface energies that control island shape and orientation in the early stages of growth.

Ag OR was observed exclusively on the Ni(100) substrate orientation. This OR has previously been referred to as the "oct-cube" OR (*OR O*) [11,12], and corresponds to: Ag(111)[1$\bar{1}$0]//Ni(100)[0$\bar{1}$1]. For a review of various systems with this OR, see [13].

The Ag ORs displayed on Ni orientations contained within the (100)-(111)-(210) triangle were referred to as "special" (*OR S*). These special ORs represent a gradual transition from *OR O*, which occurs exclusively on Ni(100), to the heterotwin OR (*OR T*) which is observed on Ni substrate orientations along the (111)-(210) line.

The (100)-(111)-(210) triangle of Ni in Fig. 1 is divided into two sub-triangles (shaded in light and medium blue); the Ag interface planes observed on Ni substrates within the light blue sub-triangle (210)-(311)-(100) have orientations that are located in the (210)-(110)-(11$\bar{1}$) (pink) sub-triangle, whereas the Ag interface planes on Ni substrates in the medium blue sub-triangle (111)-(311)-(210) have orientations that are located in the (111)-(110)-(210) (red) region of the Ag-stereographic triangle. The Ag orientations in the sub-triangle shaded in pink, may of course be re-located to the SST by changing Miller indices from (HK$\bar{L}$) to (HKL), but this obscures the simple relationship between corresponding poles. Notice that the Ag orientations in the white part of the SST do not occur at all, implying a strong texture for a Ag film grown on a perfectly random polycrystalline Ni substrate.

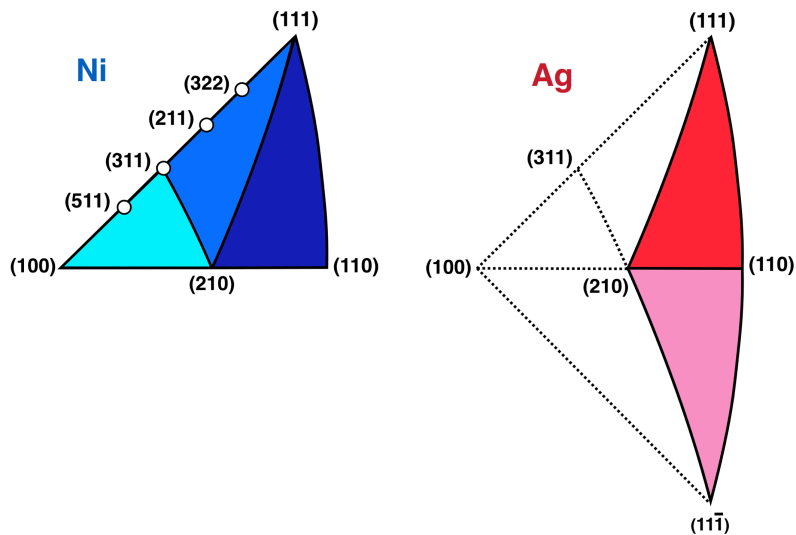

**Fig. 1**. Schematic stereographic triangles used to illustrate the experimentally determined ORs of Ag films equilibrated on Ni substrates of different orientations (hkl) [2]. The SST on the left, which identifies the Ni substrate orientations, has 3 regions distinguished by shades of blue (color on line). The dark blue triangle identifies the Ni substrate orientations on which Ag adopts the cube-on-cube OR. For this OR, Ag orientations (HKL) in the red triangle, on the right, map directly onto Ni orientations in the dark blue triangle on the left. The light and medium blue triangles identify Ni substrate orientations on which the Ag ORs are "special" (see text).

Finally, it is important to identify one additional characteristic displayed by the special ORs; the Ni[0$\bar{1}$1] and Ag[1$\bar{1}$0] close-packed directions that lie closest to the plane of the Ag/Ni interface are parallel to each other (within ~1°).

Since our study of Ag on Ni has shown *OR O* on (100) substrates to be a necessary condition for the occurrence of the variable "special" *OR S*, we decided to investigate other systems where *OR O* might be present, but where the systematic variation of *OR S* might be modified by a difference in the lattice misfit between film and substrate.

## 3. Choice of systems

Ag/Ni was chosen in our previous study because Ag and Ni have negligible mutual solubility. Such systems are convenient because the compositions of both film and substrate remain unmodified during equilibration at any temperature, i.e. no change in misfit occurs across the interface, except for minor adjustments due to differential thermal expansion effects. The Pb/Al system chosen for study here also displays negligible mutual solubility of the components [14]. However, binary FCC alloys that exhibit negligible mutual solubility are uncommon, and this criterion for system selection had to be relaxed in the case of the other system chosen for study here.

The Ag/Cu system exhibits finite mutual solubilities of the components. However, these are quite low (~1 at% Ag in Cu and ~3 at% Cu in Ag at 800 K) both experimentally [14], and as calculated by means of the Ag/Cu EAM potential [15] that has been used in the simulations performed here. When corrected for these low solubilities, the Ag/Cu lattice constant ratio is changed insignificantly from 1.131 to 1.126, compared to a ratio of 1.162 for the previously studied Ag/Ni system. In the case of Pb/Al, the ratio of lattice constants is 1.218. Thus, when taken together with the Ag/Ni system, these three systems cover a significant range of misfit, and should therefore provide insights into the role of misfit on the equilibrium ORs and interface structures that develop between films and substrates.

## 4. Computational approach

### 4.1. Molecular dynamics and molecular statics

Modeling was performed by both MD as well as molecular statics (MS) simulations using the LAMMPS code [16,17] in conjunction with EAM potentials. These potentials have been used widely in previous modeling studies of pure metals and alloys, as they include many-body effects that play an important role in metallic systems. We employed the EAM potential developed by Williams et al. [15] for simulations of Ag/Cu, the glue-type potentials (equivalent to EAM) developed by Landa et al. [18] for simulations of Pb/Al, and the potentials of Foiles et al. for Ag/Ni [10]. The EAM potentials are semi-empirical, and have typically been fitted to half a dozen, or more, materials properties. Even properties that have not explicitly been fitted to the potential are generally reproduced with acceptable values.

The computation cells for the simulations were constructed as follows. Substrates consisted of ~20,000 atoms in a rectangular crystalline slab with dimensions of approximately 10.0X10.0X2.5 nm along the *x, y* and *z* axes for Ni and Cu substrates, and approximately 11.5X11.5X2.5 nm along the *x, y* and *z* directions for Al substrates. The substrate upper *x-y* surface (which represents the interface plane with the film) was constructed so as to produce an orientation corresponding to Miller indices (hkl). The film material, consisting of ~8000 atoms, was deposited on the *x-y* surface, also in the form of a rectangular slab, in which the atoms were distributed at random. Periodic boundary conditions were applied along the *x-* and *y-*axes, but the computation cells were terminated by free surfaces in the *z-*direction.

MD simulations were performed by gradually heating the computation cells (films plus substrates) to the equilibration temperature over a period of 200 psec, holding at this temperature for periods of up to 12 nsec, and then relaxing by MS in order to remove thermal noise. The equilibration temperatures used were 800 K for Ag/Cu, 900 K for Ag/Ni, and 500 K for Pb/Al. In most cases, it was found that no further significant changes in structure occurred after equilibration times of ~4 nsec.

*4.2. Computation of orientation relationships*

Six substrate orientations were selected for study in all 3 systems: (100), (511), (311), (211), (322) and (111). In the previous simulations of the Ag/Ni system [4] only 5 of these orientations had been investigated. Here, we added simulations of Ag on Ni(211) so as to achieve a more even angular distribution of substrate planes. This particular set of substrate orientations illustrates the transition that occurs in *OR S* between (111) and (100) substrates.

Orientation relationships of the films with respect to the substrates were determined by constructing pole figures from the atomic positions in the film and substrate. This approach has been described in detail in a previous paper [4]. A brief overview is provided here.

Consider a FCC crystal. The 12 neighbors of any atom lie along <110>-type directions. When these <110> directions are redefined as unit vectors in the *x-y-z* sample coordinate frame, their projection onto the sample coordinate *x-y* interface plane yields a <110> pole figure (<110>PF) when plotted on a stereogram. The <110>PF is a convenient way of representing the orientations of both the substrate and the film, and also provides a means of determining the OR of the film with respect to the substrate.

This procedure was only applied to atoms of the film that possessed 12 neighbors of the film species, and similarly, to atoms of the substrate that had 12 neighbors of the substrate species. In effect, this process eliminates atoms that reside either at the film or substrate surfaces, as well as atoms adjacent to the film/substrate interface, which can undergo structural re-organization as a result of complexion transitions [19,20]. Discounting atoms in distorted sites focuses attention on atoms in relatively well-ordered regions with bulk-like structure, and thus reduces noise in the determination of the orientation.

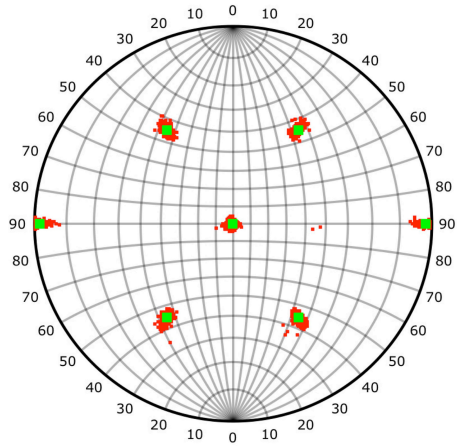

**Fig. 2.** Example to illustrate the method for determining the orientation of a film on a substrate, applied to an Ag film on a Cu(311) substrate. Each small red dot is a <110> pole computed from a given individual Ag atom and indicates the scatter of the data. Square green symbols represent the averaged Ag poles used for orientation determination (color on line).

Fig. 2 illustrates the application of this method for the case of an Ag film on Cu(311). The orientation of the film is obtained by averaging the positions of (red) points originating from the nearest neighbors of each atom in the film to yield the green symbols in the figure. Due to the local distortions of the film, the angles between the green poles are not exactly 60° and 90° as they are in an undistorted FCC crystal. To overcome the effect of these distortions, a second averaging procedure has been used to determine the Miller indices of the film interface planes. This involves determination of the film Miller indices from each set of three linearly independent (green) poles, and averaging the Miller indices thus obtained. In the case of Fig. 2, the normalized (HKL) Miller indices for the Ag film interface plane computed in this manner are (0.707 0.707 -0.002), or an orientation that lies ~0.2° from (110).

## 5. Results

In this section, the two key features of the simulations -the OR and the interface structure- are presented sequentially. The ORs are described in the form of <110> pole figures while the interface structures are shown as cross sections of the atomic structure, projected along the <110> direction that is common to the substrate and the film.

*5.1. Orientation relationships*

The ORs were determined from the simulations for all six substrate orientations in each of the three systems, by the procedures described above for constructing PFs, and are given in Table 1. However, before describing the results of Table 1 in detail, it is useful to review certain aspects of the PFs. Each PF displays the <110> poles for both the film phase (colored red) and the substrate phase (colored blue). As mentioned above, the procedure used to produce the PFs focuses on atoms with 12 neighbors in both the film and the substrate. However, in order to avoid undue clutter, our PFs only display the poles in the "northern" hemisphere. For example, for an interface plane with a general (hkl) orientation, this would reduce the number of <110> poles of both the film and the substrate from 12 to 6. However, when a <110> direction is present in the interface plane (as is the case for all substrate and film orientations reported in Table 1) then that particular <110> direction is represented by *two* <110> poles located on the outer circle of the PF at diametrically opposed locations, for a total of 7 <110> poles from each phase. This can be seen, for example, in Fig. 2, for the case of the Ag phase on a Cu(311) substrate. Also, the films generally adopt two coexisting orientations, the special orientation: *OR S*, and one of the four possible twin variants of that OR. Thus, in most of the cases displayed in Table 1, the number of <110> poles of the film phase is generally greater than 7, as will be described when a suitable example arises.

We now describe the results of Table 1, in which the rows contain the PFs for a given substrate orientation, and the columns represent the results for each of the three film-substrate materials systems, displayed in order of increasing lattice mismatch: Ag/Cu, Ag/Ni and Pb/Al.

**Table 1**
<110>PFs of substrate poles (round blue symbols) and film poles (square red symbols) for Ag/Cu, Ag/Ni [4][†], and Pb/Al. Each row of the table compares the <110>PFs obtained for a given substrate orientation. The first column provides the substrate orientation and the approximate average orientation of the film. Note that some of the figures contain a green triangle symbol representing a {111} pole about which there are 2 twin-related film orientations. Going from top to bottom, this twin plane switches from $(11\bar{1})$ to (111) of the film. Circles around red-blue pairs south of the "equator" show the systematic change in the OR. As the OR changes from *OR O* to *OR T*, the separation between the red-blue pairs decreases from 15.8° to 0° while the circles move down from the center of the pole figure as the substrate orientation moves from (100) to (111).

---

[†] Although ORs for some of the substrate orientations of Ag/Ni shown here had been obtained previously in ref [4], the present results are from new simulations, with larger numbers of film atoms, for greater reliability, as described below.

| Substrate/Film orientations (hkl)/(HKL) | Ag-on-Cu | Ag-on-Ni | Pb-on-Al | |
|---|---|---|---|---|
| (100)/(11$\bar{1}$) | | | | *OR O* $\alpha=15.8°$ |
| (511)/(33$\bar{1}$) | | | | |
| (311)/(110) | | | | *OR S* |
| (211)/(331) & its twin ~(211) (Ni-Ag & Al-Pb) | | | | |
| Ag/Cu & Ag/Ni: (322)/~(553), & its twin ~(322) Pb/Al: (322)/~(322) & an "unusual" twin ~(11 4 4) | | | | |
| Ag/Cu & Ag/Ni (111)/(111) twin & cube-on-cube Pb/Al: Pb(111)<211>/Al(111)<110> | | | | $\alpha=0°$ *OR T (OR C)* |

*5.1.1. (100) substrates*

The first row of Table 1 shows that on (100)-oriented substrates, all systems studied here display the oct-cube OR, *OR O*. The (111) orientation of the film is apparent from the triangular arrangement of square red poles, beside the substrate blue poles signifying the (100) orientation of the substrate. Note that the PF for the Ag/Cu system exhibits 3 square red poles near the center, whereas the PFs for Ag/Ni and Pb/Al systems exhibit six red poles, three of which are square and three others of which are diamond-shaped. The square red poles represent *OR O*, and the diamond shaped red poles indicate the presence of an orientation that is the twin of *OR O*. To highlight the presence of the twin relation, the twin plane normals are denoted by green triangles. This method of identifying film poles by red squares for the principal film OR, by red diamonds for the corresponding twin, and using green triangle symbols to represent a {111} pole about which there are 2 twin-related film orientations, has been used throughout Table 1. Also, if less than 3% of film atoms contributed to other new poles in a PF, they were omitted.

An ideal (100) surface will contain no surface ledges or kinks. However, in experiments performed either on single crystal substrates or on approximately (100)-oriented grains of a polycrystalline surface, it is not possible to avoid small deviations of the substrate surface from the nominal surface orientation. Thus, any realistic (100) substrate surface will display some ledges or steps (these two terms are used interchangeably). Consequently, as in the previous work on Ag/Ni [4], we have performed the present simulations on (100) substrates for both the ideal (100) substrate, with no ledges, as well as for the more realistic case, with two facing ledges running along <110> directions.

For all three systems we find that, in the presence of surface ledges, films display the ideal oct-cube OR (*OR O*), whereas when films are equilibrated on ideal planar (100) substrates, the Pb film on the Al(100) substrate undergoes a twist rotation of about 3°, but the Ag films on Cu and Ni retain the ideal oct-cube OR. Thus, the presence of ledges on the substrate leads to an alignment of the film in Pb/Al. All the pole figures for (100) substrates displayed in Table 1 are for the case of (100) substrates with ledges. However, the <110>PF for the case of the ledge-free Pb/Al(100) film, which shows a twist rotation of about 3°, is displayed in Fig. 3.

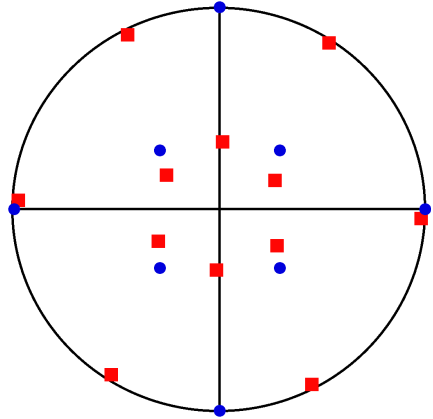

**Fig. 3.** <110>PF of Al(100) substrate poles (blue round symbols) and Pb(111) film poles (red square symbols) for the case of a ledge-free substrate. Comparison of this figure with the corresponding one for an Al(100) substrate with ledges in Table 1 shows that the <110> directions in the Pb film in this ledge-free case are not exactly parallel to <110> directions in the substrate, but are rotated by about 3° (see text).

*5.1.2. (511) substrates*

For the (511) substrates, the pole figures for all three systems are essentially identical, with a principal film OR close to $(33\bar{1})$. Twin-related poles (red diamonds) of the principal OR appear in two of the systems (Ag/Cu and Ag/Ni).

*5.1.3. (311) substrates*

Again the results for all three systems are essentially identical, and are all consistent with a film interface orientation of (110). However, in this case, the PFs do not display any twin-related poles.

*5.1.4. (211) substrates*

For (211) substrates, the films adopt an orientation close to (331) (red squares) in all three systems. It is slightly away from (331) in Pb/Al (see the separation between blue and red poles inside the marked circles). The fact that the red diamonds denoting the twin almost overlap the blue (211) substrate poles indicates that the orientation of the twin in the film is close to the (211) orientation of the substrate.

*5.1.5. (322) substrates*

In our previous study of Ag films equilibrated on 12 different Ni substrates [4], it was found that the (322) orientation was one of only two cases where the Ag film displayed four different orientations. As a result, there was some uncertainty in the determination of film ORs. Attempts to reduce the data overlap by longer MD runs (up to 12 nsec) were unsuccessful. Consequently, the inferred <110> poles of the Ag deposit on Ni(322) reported in the previous study [4] were only approximate. In the present study, another approach for reducing scatter was attempted, namely, doubling the film thickness on all substrates from ~4000 to ~8000 atoms. This approach was successful in reducing the number of ORs of Ag/Ni(322) to two. So, the approach of increasing film thickness was extended to all the other simulations.

As shown in the fifth row of Table 1, on (322) substrates, the films of two systems adopt a (553) interface plane orientation as well as the twin of this orientation, which is close to (322) by rotation about the (111) pole (green triangle). However, in the Pb/Al system, the two coexisting ORs are close to *OR C* and to an unusual twin. This twin corresponds to rotation about the ($\bar{1}$11) pole tilted by ~80° from the substrate normal (see green triangle), rather than the (111) pole that is almost parallel (tilted by ~11°) to the substrate normal, as shown for the Ag/Cu and Ag/Ni systems. Note also that for Pb/Al the small misorientation between blue and red poles in the lower hemisphere of the PF is in the opposite sense to that for the other two systems.

*5.1.6. (111) substrates*

Here the interface orientations of the films for Ag/Cu and Ag/Ni are identical, and correspond to a combination of *OR T* and *OR C*. As for (100) surfaces, two facing ledges were added to the (111) substrates. This prevents an in-plane rotation of the Ag films on the Ni and Cu substrates, but not for Pb on Al. The corresponding figure for Pb/Al shows that the Pb film also displays a {111} interface orientation, and also exhibits twinning, as evidenced by the hexagonal patterns of Pb<110> poles. On the outer edge of the pole figure the red and blue poles are separated by 30° indicating an in-plane twist between Pb and Al. Addition of <110> steps to the Al(111) substrate plane in this case does not change the OR of Pb, and it is found that it is the Pb<211> steps that align with Al<110> steps.

*5.1.7. Summary of Table 1*

The ORs illustrated in Table 1 follow a systematic pattern that is most apparent in the first two columns. Consider the red-blue pairs of poles south of the equator (marked by circles). Going up a column from (111) to (100) substrates, the separation of these poles increases gradually in a pattern that is characteristic of *OR S*. This is a visual indication that *OR S* is maintained throughout the table, gradually changing from *OR T* at the bottom to *OR O* at the top. In addition, almost all films exhibit a second orientation of the red poles, which is twin-related to the first orientation. The only two exceptions to this systematic pattern of ORs are seen in the bottom two panels of the Pb/Al column. Like the other systems, the Pb films form in two orientations that are twin-related to each other. However, for Pb/Al(322), the twin plane does not lie near the interface, but rather makes a ~80° angle with it. And for Pb/Al(111), both the film and its twin are twisted by 30° to the substrate about the substrate normal.

Overall, Table 1 shows a pattern of ORs that is essentially independent of mismatch for 16 of the 18 cases illustrated, i.e. with the exceptions of Pb on Al(322) and on Al(111).

*5.2. Interface structure*

While the ORs are found to be essentially identical for all three systems, the effect of the mismatch is apparent in the interface structure. This is illustrated in Fig. 4 with cross sections of a few representative interfaces in different systems grown on various substrates.

The atoms in the film are colored using the common neighbor analysis [21]. Atoms shown in yellow are FCC-coordinated, atoms in blue are HCP-coordinated, while atoms in green are in an environment that cannot be classified. The green colored atoms are mostly located either at the film surface, where atoms lack full coordination, or at the film interface with the substrate, where the environment straddles two different structures, and cannot therefore be readily identified. Also, these green colored atoms are excluded from the pole figures, and do not contribute to the OR. Substrate atoms are colored red for Cu, orange for Ni and purple for Al.

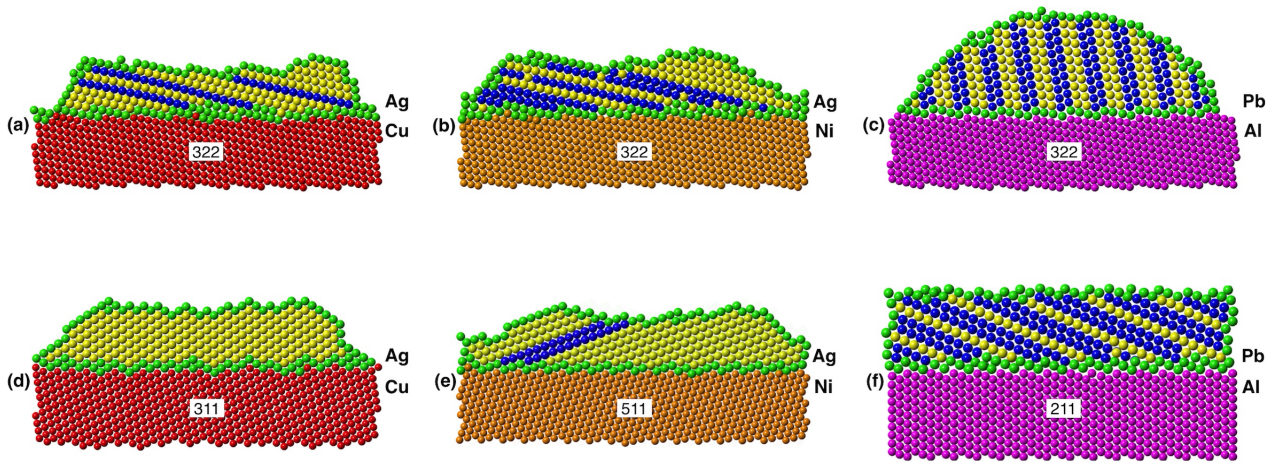

**Fig. 4**. Cross sections of films equilibrated on different substrates for the three systems Ag/Cu, Ag/Ni and Pb/Al with a mismatch of 12.6, 16.2 and 21.8%, respectively. The viewing direction is the common <110> direction that lies in the interface. Atoms in the films have been colored by the common neighbor analysis, with FCC in yellow, HCP in blue and "Unknown" in green (color online). The blue segments indicate the presence of stacking faults. Close-packed film planes parallel to the substrate terraces meet edge-to-edge in the interface. Faults are parallel to the terrace planes except in c). The first row (a-c) compares films on (322) substrates for the three systems. Films on (311) substrates (d) are free of stacking faults. Faults in Pb films are periodic in c) and f).

A striking feature in most of the cross sections is the presence of stacking faults, as indicated by the bands of blue atoms, which belong to {111} planes that lie parallel to the viewing direction. As will be shown below, these faults account for the twin orientations in the pole figures.

A second striking feature common to all simulated interfaces is that the planes parallel to the substrate terraces in both phases meet edge-to-edge at the steps in a one-to-one correspondence. Surprisingly, this is the case for all three systems, even though the plane spacing is significantly larger in the film than in the substrate. Even in Pb/Al, despite a 21.8% mismatch between plane spacings, these planes still meet edge-to-edge, without disregistry. Across each terrace, the close-packed planes facing each other in film and substrate are parallel. But at a distance from the interface they deviate by a small angle that depends on the orientation of the substrate and the mismatch.

A more detailed inspection of the stacking faults in films grown on (322) substrates (Fig. 4a-c) shows that there are 3 faults in Ag/Cu, 5 faults in Ag/Ni and 9 faults in Pb/Al. However, in Pb/Al, they lie on the ($\bar{1}$11) plane, at 70.5° to the interface terrace planes, an exception to all the other interfaces, where the stacking faults are parallel to the interface terraces.

The unusual inclination of the fault plane on (511) substrates apparent in Fig. 4e is due to the different orientation of the terrace plane on substrates near (100). This is best seen from the orientation of the terraces in the step+terrace structure at the lower substrate surface of the cross sections. Note that on the (311) substrate (Fig. 4d), the interface does not display a clear step+terrace structure. Films on this substrate are also free of stacking faults in all three systems (cf. Table 1), although planes clearly meet edge-to-edge across the interface.

Two of the films were found to exhibit periodic faults. Fig. 4f shows a periodic array of faults parallel to the terrace planes for (211) substrates in Pb/Al. With a fault on every third plane, this is the 9R structure. Stacking faults lead to nearest-neighbor bonds that are twin-related to the unfaulted crystal. This is perhaps most readily apparent in the Pb film in Fig. 4c, where two blue layers are separated by two yellow layers, forming essentially a periodic 12R structure. If we treat each color segment as a two-layer slice of a more extended structure, it can be seen that the two are twins of each other. Thus, poles marked as a twin of the film in all the <110> pole figures of Table 1 do in fact arise from the presence of stacking faults. Note that pole figures measure the orientation of nearest-neighbor bonds, but not their spatial distribution. This makes a twin indistinguishable from an array of separate intrinsic or extrinsic stacking faults.

The same characteristic structure and stacking faults (although varying in density) were found for all orientations and in all three systems. As will be described in the Discussion, these faults play an important role in accommodating the mismatch between the substrate and the film.

## 6. Discussion

The pattern of heteroepitaxial growth presented above can be understood in the framework of transformation strains. In this Discussion, it will be shown that the difference between *OR C* and *OR S* can be explained by transformation strains for two different correspondence relationships between the substrate lattice and the film lattice. The following sections will first outline these two lattice correspondence relationships, then apply a previously developed model [3] to all of the data in Table 1, and finally explore the role of defects in the simulated interfaces.

*6.1. Lattice correspondence*

As illustrated in Figure 1, the pattern of heteroepitaxial growth on different substrate orientations comprises two domains. On Ni substrates in the right half of the SST (dark blue) Ag films grow in *OR C*, while substrates in the left half of the SST lead to *OR S*. Whereas *OR C* is fixed, *OR S* is continuously variable. These two domains are characterized by two distinct correspondence relationships: the cube-cube correspondence for *OR C* and the heterotwin correspondence for *OR S*.

A lattice correspondence is a set of vectors or planes in the parent lattice, which after transformation become a corresponding set of vectors or planes in the product lattice. The transformation can then be described as a homogeneous deformation of the parent into the product lattice (e.g. [22-24]) In Fig. 1, the two lattice correspondences can be visualized as mapping regions of orientation space from Ni to Ag. In the SST, the dark blue triangle of Ni maps directly to the red triangle in Ag, resulting in *OR C*. This can be written as "$(hkl)_{Ni}$ corresponds to $(HKL)_{Ag}$", denoted $(hkl)_{Ni} \triangleq (HKL)_{Ag}$. Choosing corresponding poles at the corners of the dark blue and red triangles, we have:

$$(111)_{Ni} \triangleq (111)_{Ag}$$
$$(110)_{Ni} \triangleq (110)_{Ag}$$
$$(210)_{Ni} \triangleq (210)_{Ag}$$

More generally, this relationship can be written in terms of the correspondence matrix **M**

$$\begin{pmatrix} H \\ K \\ L \end{pmatrix} = \mathbf{M} \begin{pmatrix} h \\ k \\ l \end{pmatrix} \quad (1)$$

For the cube-cube correspondence, **M** is simply the identity matrix **I**. However, for the heterotwin correspondence, **M** takes a more complex form. In Fig. 1, this can be seen from the mapping of the left triangle in Ni (bounded by (111), (210) and (100)) to the larger right triangle in Ag

(bounded by (111), (210) and (11$\bar{1}$)). Corresponding poles at the corners of these triangles are then:

$$(111)_{Ni} \triangleq (111)_{Ag}$$
$$(210)_{Ni} \triangleq (210)_{Ag}$$
$$(100)_{Ni} \triangleq (11\bar{1})_{Ag}$$

leading to a correspondence matrix

$$\mathbf{M} = \frac{1}{2}\begin{pmatrix} 1 & 2 & \bar{1} \\ 1 & 0 & 1 \\ \bar{1} & 2 & 1 \end{pmatrix} \quad (2)$$

It is the difference between the angular ranges (54.7° and 70.5°) of these two corresponding triangles that causes the variation in *OR S*.

To explain why the systematic variation in *OR S* is mostly identical for the three systems investigated here, Fig. 5 illustrates the relationship between the two lattices, using the Ag/Ni system as an example. The Ni lattice is shown in blue and the Ag lattice (16.2% larger) in red. Fig. 5a highlights the heterotwin relationship between the two FCC unit cells as a 180° rotation around the axis normal to their common {111} plane, highlighted by shaded triangles. Within this heterotwin plane, the crystals maintain a common close-packed <110> direction, marked by arrows. By projecting the corresponding lattices along this common <110> direction, as shown in Figs. 5b and c, we reduce the analysis of the ORs to a two-dimensional problem. Note that the 16.2% mismatch in the 3$^{rd}$ dimension, along the common <110> direction, remains fixed for all ORs and all interfaces considered.

In Fig. 5b, the two lattices meet in the heterotwin plane (seen edge-on), and two corresponding unit cells are outlined. To visualize the deformation that transforms one lattice into the other, a dashed Ni cell has been superimposed directly on the Ag cell. It is apparent that the deformation can be described by a shear, *s*, parallel to the twin plane, plus a uniform expansion by the mismatch $\rho(=a_{film}/a_{substrate})=1.162$. This can be written as $\rho\mathbf{S}$, where $\mathbf{S}$ is a simple shear of magnitude *s* on the common {111} plane, with

$$s = \tfrac{1}{4}\sqrt{2} = 0.35 \quad (3)$$

While this leaves the common {111} plane stretched but not rotated, all other corresponding planes, (hkl)$_{Ni}\triangleq$(HKL)$_{Ag}$, will be rotated relative to each other during the transformation. To illustrate this rotation directly, the two twin-related lattices are shown side by side in Fig. 5c, with some corresponding planes in Ni and Ag outlined and labeled. These corresponding planes were obtained from Eq. (1). Note that due to the inverse relationship between planes and directions, corresponding directions are obtained from [UVW]$_{Ag}=\mathbf{M}^{-1T}$[uvw]$_{Ni}$ [e.g. 25].

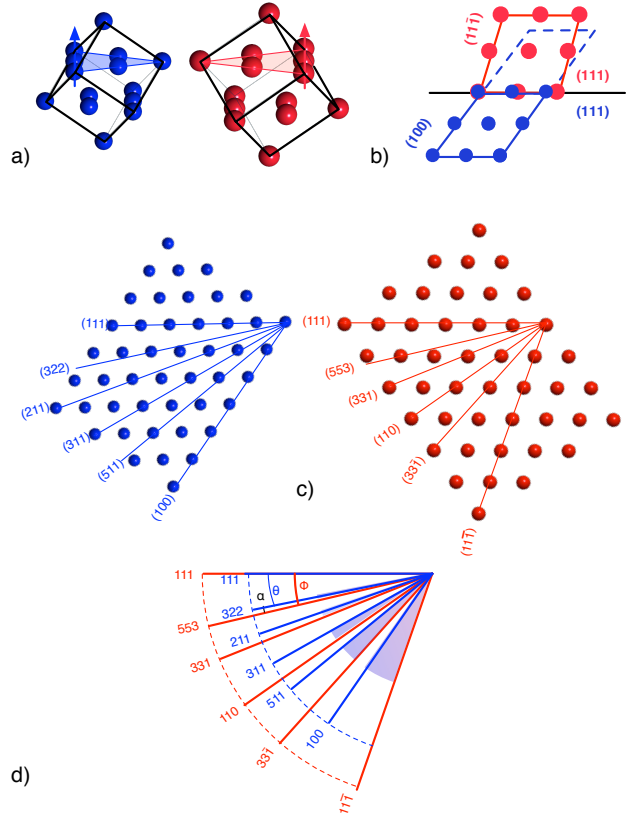

**Fig. 5**. Illustration of heterotwin lattice correspondence with Ag/Ni as an example, showing Ni (blue) and Ag (red) expanded by 16.2% relative to the Ni. In a) the two unit cells are shown in a perspective view, highlighting the common twin plane. The arrows mark the common $[0\bar{1}1]_{Ni}//[1\bar{1}0]_{Ag}$ direction, which is the viewing direction in the 2D projections shown in the rest of the figure. In b) the two lattices meet in their common (111) plane. Corresponding unit cells are outlined in the two lattices. The deformation that transforms the dashed Ni cell to the red Ag cell can be visualized as a shear of magnitude $s$=0.35 parallel to the (111) plane plus a uniform expansion of 16.2% (the mismatch between Ni and Ag). In c) corresponding planes in the two lattices are outlined and labeled. Note that the angular relationships between corresponding planes do not depend on the mismatch because the angles between lattice planes do not change under a uniform expansion or contraction. To illustrate the increasing angle α between corresponding planes, the two fans of planes are superimposed in d), with α=ϕ−θ marked by shaded wedges, increasing from 0 to 15.8° as substrate orientations go from (111) to (100).

As noted before [3,4], the fans of corresponding planes in Fig. 5c cover different angular ranges, 54.7° in Ni and 70.5° in Ag. Thus, when (111) planes are parallel, all the other corresponding planes diverge by an angle α that increases from 0° at $(111)_{Ni}\triangleq(111)_{Ag}$ to 15.8° at $(100)_{Ni}\triangleq(11\bar{1})_{Ag}$. For two corresponding planes to meet face-to-face in the interface, they must be parallel, i.e. unrotated under the transformation. To eliminate the angular gap caused by the deformation $\rho\mathbf{S}$ requires a lattice rotation **R** through an angle α. The angular gap is illustrated in Fig. 5d for the corresponding planes examined in this study. The total transformation **R**$\rho$**S** then ensures that the interface plane is stretched but not rotated by the transformation. This means that the interface is an unrotated plane, i.e. in addition to the common close-packed directions, it contains another eigenvector (a direction that is stretched but not rotated during the transformation). The magnitude of the stretch is described by the eigenvalue λ, and the strain is given by $\varepsilon=\lambda-1$.

Figures 5c,d illustrate why the ORs for most of the interfaces investigated in this study are independent of mismatch. Imagine changing the mismatch by uniformly expanding or contracting the Ag lattice in Fig. 5c. This will not affect the angular relationship between corresponding planes, i.e. a given substrate surface $(hkl)_{substrate}$ will be met face-to-face by the corresponding plane $(HKL)_{film}$ in the film, regardless of mismatch. However, the *magnitude* of the strain, and hence the strain energy in the interface depends directly on the mismatch. If, for a given substrate surface, the interfacial strain for the heterotwin correspondence is too large, the crystal may choose a different lattice correspondence, e.g. the cube-cube correspondence.

To evaluate the magnitude of the strain, $\varepsilon=\lambda-1$, in an interface, we use the approach described in our previous analysis of the Ag/Ni system [3] which compared a linear, a 1D, and a 2D approach. Overall, the 2D model, based on transformation strains, provided the best fit to the data obtained for the Ag/Ni system. In the following section we show that the same approach can explain why the above behavior appears to be independent of the ratio of lattice parameters, by applying it to all the results summarized in Table 1.

*6.2. Application of the 2D model to systems with different mismatch*

The transformation matrix for the three systems can be written **R**$\rho$**S**, where **S** is a simple shear of magnitude 0.35 on the common (111) plane of the heterotwin correspondence, $\rho$ is the ratio of lattice constants and ranges from 1.126 for Ag/Cu to 1.218 for Pb/Al, and **R** is the lattice rotation needed to fill the angular gap $\alpha=\phi-\theta$ between (hkl) and the corresponding (HKL). The strain in a given interface can be obtained from the eigenvalue of the eigenvector contained in the interface.

An explicit solution for the eigenvector equation of the transformation **R**$\rho$**S** was given in ref. [3]:

$$\lambda_{1,2} = \rho/2 \, (B \pm \sqrt{B^2-4}) \quad = \quad 1+\varepsilon_{1,2} \tag{4}$$

where the two solutions $\lambda_{1,2}$ (or $\varepsilon_{1,2}$) correspond to the positive and negative values of the square root, and $B \equiv 2\cos\alpha + s\sin\alpha$ depends on the shear *s* and the rotation angle α. A key feature of Eq. 4 is the fact that the eigenvalues $\lambda_1$ and $\lambda_2$ are directly proportional to $\rho$. In contrast, the corresponding directions (eigenvectors) are independent of $\rho$ [3]. This can be seen directly in Fig. 5c where the two fans of corresponding planes are independent of the factor $\rho$ that describes the relative magnification of the two lattices. It is also consistent with the observation that the OR and interface plane show mostly identical behavior (i.e., the same trend of rotation) for different substrate surfaces in all three systems investigated (the two exceptions to this pattern will be discussed later). This simply means that while the *directions* of the strain depend only on the substrate surface, the *magnitude* of the strain is proportional to the mismatch.

*6.3. Variation of interface strain with mismatch*

To illustrate the dependence of the strain, $\varepsilon$, on the mismatch, Fig. 6 presents a plot of Eq. 4 for the case of α=15.8°, i.e. the oct-cube OR (*OR O*). This figure represents $\varepsilon$ as a function of the lattice parameter ratio $\rho$, showing $\varepsilon_1$ in orange and $\varepsilon_2$ in purple. For comparison with the cube-on-cube OR, the strain for *OR C* is shown as a dashed green line, to indicate an extension by a factor $\rho$ (i.e. $\varepsilon_{OR\ C} = \rho-1$). It is apparent that the orange, green and purple lines each have a range of mismatch $\rho$ where the magnitude of strain is smaller than that for the other two lines. The cross-over between their ranges occurs at $\varepsilon=\pm7.2\%$, indicated by a yellow band. The three systems investigated in the present study are indicated by light blue vertical lines crossing the x-axis and are seen to fall clearly into the range where the strain indicated by the orange line is smaller than the other two. This means that on a (100) substrate, films grow in (111) orientation (*OR O*).

If we reverse the role of substrate and film, the ratio of lattice parameters is inverted, and the minimum strain changes to the (left) purple line ($\varepsilon_2$). In this case, the substrate plane is (111), and the film grows in (100) orientation, again as in *OR O*, but in reverse. However, we are not aware of any experimental observations of such a case[‡], and no simulations have confirmed this case either. For example, for Ni on Ag (111) (for which $\rho=0.86$), simulations find a preference for Ni (111) [11], instead of the Ni (100) suggested by the present strain analysis.

Fig. 6 shows that in a choice between *OR C* and *OR O* on a (100) substrate, *OR C* has the lowest absolute value of strain when $\rho$ is between 1.07 and 0.93, while outside of this range, *OR O* has the lower absolute strain. This supports the conclusion of Hsiao et al. [28] that *OR O* requires a minimum misfit strain of about -8%. However, the present work has investigated not just the interfaces (unrotated planes) in *OR O*, but the whole range of *OR S* from (111) to (100) substrates.

---

[‡] Although observations of (100) FCC metal films on (111) diamond cubic substrates, such as Al on Si and Ge [26,27] could be counted as such a case, the difference between FCC and diamond cubic crystal structures would complicate the simple strain analysis.

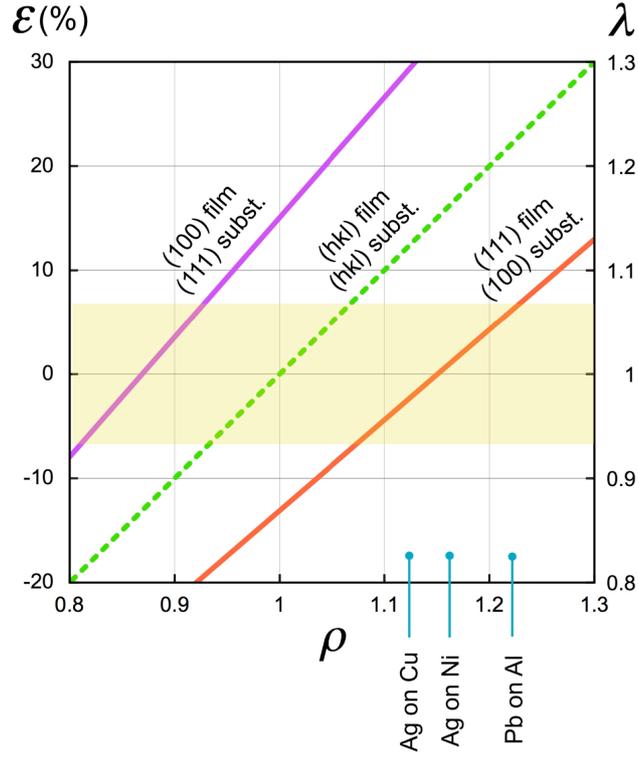

**Fig. 6.** Graph showing interface strain $\varepsilon$ (eigenvalues $\lambda$) for *OR O* ($\alpha$=15.8° i.e. *B*=2.02) (solid orange and purple lines) and *OR C* (dashed green line), as a function of lattice mismatch, $\rho$. The yellow horizontal band denotes the range of $\varepsilon = \pm 7.2\%$. Within this range of strain, each curve occupies a corresponding range of lattice mismatch, $1.07<\rho<1.23$ for the orange line, $0.81<\rho<0.93$ for the purple line, and $0.93<\rho<1.07$ for the green line.

*6.4. Variation of interface strain with substrate orientation*

To see how the strain varies with substrate orientation we plot $\varepsilon$ versus $\alpha$ as given by Eq. 4, setting $\rho$ to its values for Ag/Cu, Ag/Ni and Pb/Al. This plot is shown in Fig. 7. The $\alpha$-axis covers the range from 0 to 15.8°. For reference, the substrate orientations explored in this study are indicated by light blue vertical lines. These orientations are also visible as the blue fans of plane traces in Figs. 5c and d. The dotted horizontal lines are the strains for *OR C*, where all directions undergo an extension by a factor $\rho$ (i.e. $\varepsilon_{OR\ C} = \rho-1$). To read this graph, take for example a (311) substrate, where the strains for the 3 systems range from about -5% for Ag/Cu to about +5% for Pb/Al. By comparison, the strains for *OR C* (dashed lines) range from 12.6% for Ag/Cu to 21.8% for Pb/Al.

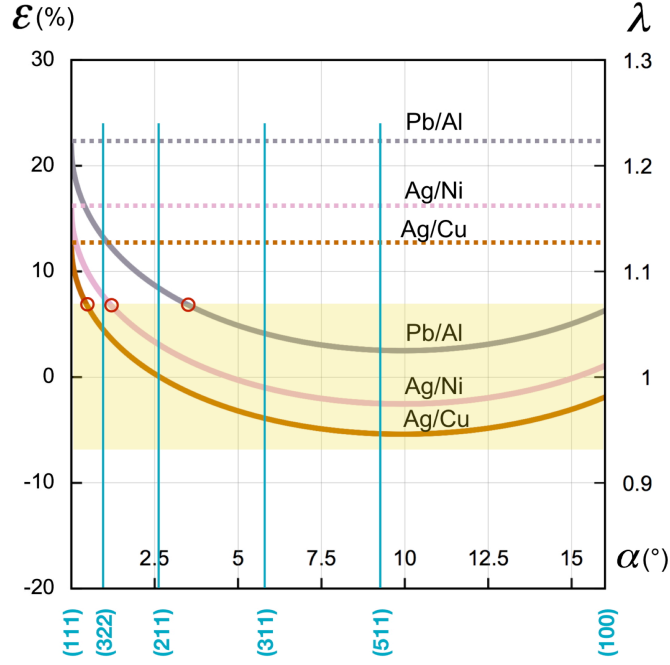

**Fig. 7.** Comparison of strains $\varepsilon$ (eigenvalues $\lambda$) for *OR S* (solid curves) and *OR C* (dotted horizontal lines) for the three systems considered. The shaded horizontal band indicates $\varepsilon=\pm7.2\%$, and its intersection with the solid curves (*OR S*) is marked by circles. Note that the strain for *OR S* lies within the $\pm7.2\%$ band for most of the interfaces, while the strain for *OR C* is significantly larger, except at (111) substrates, where *OR S* and *OR C* have the same strain.

For most of the α-range, all 3 systems lie within the ±7.2% yellow band. A notable exception is the (111) substrate, where the strain for *OR S* is identical to that for *OR C*. Apart from the (111) substrates, (322) and (211) substrates for Pb/Al also clearly lie outside the ±7.2% band. As shown below, this is reflected in the interface structures and ORs for these systems.

*6.5. Interface structure and edge-to-edge plane matching*

The systematic pattern of ORs described in Table 1 closely matches the predictions of the 2D model [3]. This match rules out alternative models, specifically the 1D model, which is based on edge-to-edge matching [24,29] and related to the topological approach [30]. Yet, as shown in the Results, the simulations show clearly that edge-to-edge plane matching was maintained for all interfaces. These two observations point to an apparent contradiction within the simulations: most of the ORs agree with the 2D model, while the interface structure supports the 1D model.

The difference between the ORs for the two models can be visualized by reference to Fig. 5d. Corresponding planes in the substrate and film lie at angles $\theta$ and $\phi$, respectively, to the {111} plane of the twin correspondence. The 2D model requires that the film be rotated by $\alpha_{2D} = \phi - \theta$ to make corresponding planes parallel in the interface. By comparison, the 1D model requires that the terrace planes meet edge-to-edge in the interface, demanding that $\alpha_{1D} =$ $sin^{-1}(\rho \sin\theta) - \theta$ for {111} terraces. Unlike the 2D model, this rotation depends on the lattice mismatch $\rho$.

For the example of the Ag/Ni(322) interface in Fig. 4b, Table 1 shows that the corresponding planes facing each other in the interface are $(322)_{Ni} \triangleq (553)_{Ag}$, in good agreement with the 2D model, which requires a rotation of $\alpha_{2D}=0.9°$ to close the angular gap between these two corresponding planes. In contrast, the 1D model requires $\alpha_{1D}=1.8°$. Fig. 4b shows that the (111) planes parallel to the terraces clearly meet edge-to-edge in the interface, as seen by the matching of these planes' edges at all the steps.

Close examination of all the other interfaces simulated in Table 1 showed the same apparent contradiction in most cases. While the ORs followed the 2D model, Fig. 4 illustrates that all interfaces displayed edge-to-edge matching for the terrace planes, implying the OR of the 1D model.

The key to this paradox lies in the stacking faults. The simulations show that the films are distorted to accommodate the mismatch with the substrate in the interface. The stresses due to this distortion can be relieved by dislocation shear. For partial dislocations, this will lead to stacking faults. The magnitude of the mismatch in the interface is given by the eigenvector in the interface, which is stretched but not rotated. For *OR C*, the stretch is simply equal to $\rho$, while for *OR S*, the stretch is the eigenvalue $\lambda$, given by Eq. 4 and shown in Fig. 7.

For low interfacial strain $\varepsilon=\lambda-1$, the stress resolved onto the terrace plane may be insufficient to generate faults. For example, for Ag on Cu(211), the strain is close to zero and there are no stacking faults (see Table 1). However, most of the other films exhibit faults, which are generally parallel to the terrace planes of the step+terrace structure of the interface. For most of the interfaces in Fig. 4, the strains are within the range of ~7.5%, except for Pb/Al, where the strain is 13.4% on the (322) substrate and 8.5% on the (211) substrate. This effect of the mismatch is reflected in unusual structures with a high density of periodic stacking faults on different planes (Fig. 4c and f).

Fig. 4e shows that for (511) substrates, the fault plane is inclined in the opposite sense to that of the other films. This is due to a change in the terrace planes as the substrate approaches the (100) orientation and can be understood by reference to Table 1. While in rows 1 and 2 of the table, the stacking faults lie on the $(11\bar{1})$ plane, in rows 4-6, they lie on the (111) plane. These are the terrace planes in the step+terrace description of the film, namely the (100) and (111) planes of the substrate, respectively. Surfaces vicinal to (100) are made of long (100) terraces and short (111) steps. Conversely, surfaces vicinal to (111) are made of long (111) terraces and short (100) steps. In the microfacet decomposition scheme of the TLK (Terrace-Ledge-Kink) model of surfaces [31], terraces and steps (or ledges) exchange roles when they are equal in length, which occurs at the (311) substrate orientation, corresponding to the (110) plane in the film. This interchange is reflected in the change in orientation of the stacking faults (or twin), as seen in Table 1 and in the interface cross sections displayed in Fig. 4.

Table 1 shows that for (311) substrates, films in all 3 systems are free of stacking faults (or twins). In Fig. 4d, this can be seen directly for the interface in the Ag/Cu system. The interfacial strain of -3.8% is accommodated by elastic distortion, allowing close-packed planes to meet edge-to-edge. The same behavior was found for the other two systems, where the interfacial strains are -1% and +4.2%, respectively. Hence interfaces on (311) substrates are coherent, whereas all the other interfaces are semicoherent, with a well-defined set of dislocations, as indicated by the stacking faults. Apparently, such large elastic distortions are possible because of the small size of the computational cell and because the comparable thickness of film and substrate distributes the strain between them. Only for simulations with thicker films (not shown here) did stacking faults emerge, at times on both types of {111} plane, and occasionally including even a fault on the other two {111} planes, at an angle to the common <110> direction. This behavior is typical of heteroepitaxial growth when the critical film thickness is exceeded [1].

Despite their absence on (311) substrates, stacking faults are clearly an essential feature of the simulated interface structure, becoming more important for larger interfacial strains, such as Pb/Al (211) ($\varepsilon$=8.5%) and (322) ($\varepsilon$=13.4%). As seen in Fig. 4c and f, in both of these cases, the stacking faults are periodic, forming a 9R structure on (211) substrates and a 12R structure on (322) substrates. Whether the occurrence of these periodic structures is coincidental, due to the specific magnitude of the interfacial strain, or indicative of metastable structures stabilized by the interface [32-34] remains to be seen. However, there are important differences between the two cases of periodic faults in Pb/Al. Not only is the periodicity of the faults different (a fault for every step on (322) substrates and a fault for every other step on (211) substrates), but the fault planes are different as well, as can be seen from their inclination to the interface in Fig. 4. For (211) substrates, the faults are parallel to the terrace planes, while for (322) substrates, they are steeply inclined. Below, we examine these two cases in more detail.

*6.5.1. Pb on Al(211)*

The behavior on a (211) substrate, shown in Fig. 4f, is in line with that in the other systems, although due to the larger strain, edge-to-edge matching can no longer be enforced by elastic distortion. Instead, the 8.5% interfacial strain is fully relieved by periodic stacking faults spaced 3 planes apart. A Shockley partial on every third plane amounts to a shear of s=(1/6

<112>) / (3·1/3 <111>) = 0.24 on the terrace plane. The steps generated by the faults, where they meet the interface, produce a step+terrace structure that changes the average orientation of the interface plane in the film by an angle $\omega \cong s \sin^2 \phi = 1.9°$. Here $\phi$ is the angle between the (111) terrace plane and the (331) interface plane in the film. To return the stepped plane to its position parallel to the substrate surface, the film must be rotated by 1.9° in the opposite sense. This accounts for the ~2° difference between the 2D model and the interface observed in Fig. 4f. Note that the same shear of s=0.24 also causes a compression $\varepsilon \cong s \sin \phi \cos \phi$ along the interface plane that eliminates the 8.5% expansion of the eigenvector. The phase transformation followed by the shear due to the stacking faults leaves the interface strain-free overall. This implies that edge-to-edge matching, when accommodated by a lattice-invariant shear (the stacking faults) produces an invariant line strain [24].

*6.5.2. Pb on Al(322)*

As shown in Fig. 4c, the periodic faults on Pb films grown on Al (322) substrates lie on a plane that is steeply inclined to the interface, and the film follows *OR C* rather than *OR S*, like all the other cases. This can be understood as a result of the mismatch.

The strain along the interface, which is 8.5% for the (211) substrate, increases to 13.4% for the (322) substrate. To accommodate this strain by stacking faults would require a shear of s=0.64, almost 3 times that for the (211) substrate. While the interface strain in films on (211) substrates is accommodated by a Shockley partial dislocation on every 3rd terrace plane (s=0.24), the interface strain on (322) substrates could be eliminated by placing a Shockley partial on every plane (s=0.71). Since this describes the twinning shear, such a fault sequence would amount to detwinning the heterotwin, returning it to the cube orientation. This is why Pb prefers *OR C* on (322) substrates. By comparison, the smaller interface strains for Ag/Ni and Ag/Cu, (7.6 and 4.4%, respectively) can be accommodated by shears of magnitude 0.37 and 0.21, respectively. These shears can be generated by a fault on every other plane (s=0.35) or every third plane (s=0.24), i.e. ~5 faults for Ag/Ni and ~3 faults for Ag/Cu, as observed in Figs. 4a,b. Since the larger interface strain for (322) substrates in Pb/Al cannot be accommodated for *OR S*, the system utilizes different faults in *OR C* to achieve edge-to-edge matching of the terrace planes. This unusual mechanism is described below.

In Fig. 4c, the periodic faults are seen to be steeply inclined to the substrate. Note that two blue layers are separated by two yellow layers. The direction normal to the (322) substrate is (322)$_{Pb}$ for the yellow segment and (11 4 4)$_{Pb}$ for the blue segment; this means that the two segments follow *OR C* and its twin. The unusual feature is that unlike *OR S*, the twin plane is not the Pb (111) plane that is parallel to the substrate terraces, but the ($\bar{1}$11) plane inclined at 70.5° to the terrace plane. In Table 1 this difference is visible from the position of the green triangle indicating the normal to the twin plane.

These features can be understood by noting that in Fig. 4c, the Pb (111) planes that are parallel to the substrate terraces are perfectly registered across the interface despite the 21.8% difference in their spacing. This is made possible by the stacking faults associated with the steps. Whenever a stacking fault intersects a surface, it will leave a step whose height is a fraction of a perfect step (1/3 or 2/3 for a Shockley partial dislocation). Thus the stacking faults attached to the steps can reduce the step height by 1/3, or 33%. Since this overcompensates the 21.8% difference in plane spacing, it requires a small counter-rotation of the Pb lattice by ~2°, in the opposite sense to the lattice rotations observed for all the other interfaces. Meanwhile the terrace length of 4Pb/5Al spacings is in the ratio of 1.25, close to $\rho$=1.218, and may be considered a "magic size" [35,36]. This means that to accommodate the 21.8% mismatch, each terrace has exactly one extra Al plane (perfect dislocation), while each step has exactly one Shockley partial dislocation, constituting a disconnection [30].

More generally, it is evident that the interface must be unrotated and unextended overall. To avoid accumulating long-range elastic strains, any extension $\varepsilon$ caused by the transformation **R$\rho$S** must be accommodated by a lattice-invariant deformation, i.e. a series of dislocations and associated steps. Whether these dislocations are generated by shear or climb is immaterial to the structure of the interface, although this has important implications for its mechanism of motion.

*6.6. Films on (111) substrates*

The only other deviation from the general pattern of ORs in Table 1 was found for Pb films on Al (111) substrates. Simulations show that Pb grows in a (111) orientation, but rotated by 30° about the common [111] direction. This is a new OR that does not fit the pattern observed for all the other cases studied here, and it highlights the limit of a model based solely on strain.

Clearly, any strain of more than a few percent cannot be accommodated by elastic deformation but must involve some form of atomic rearrangement at the interface. For the Ag/Ni(111) interface this rearrangement takes the form of misfit dislocations (e.g. [12]) that accommodate the 16.2% mismatch. In surface science this atomic rearrangement within monolayers on rigid substrates has been discussed in the framework of static displacement waves [37] or reconstructions [38]. As displayed in Fig. 3, our simulations show that a Pb film prefers a 3° rotation on a perfectly flat Al(100) substrate that is free of steps. Similar small rotations were found for Ag(111) on step-free Ni(111) surfaces [4,38], as well as on step-free Cu(111) surfaces as observed in the present simulations. Note that such rotations cause the <110> directions that usually remain parallel for *OR S* and *OR C* to spread apart and thus break the common 2-fold symmetry of the bicrystal. Simulations clearly indicate that the associated reconstruction lowers the interface energy.

Structurally, a reconstruction can be viewed as a network of dislocations. For example an in-plane twist rotation of $\omega$ could be described as a network of screw dislocations with Burgers vector *b* and spacing *L* where $b/L= \tan\omega$. For a rotation of $\omega=3°$, the dislocation spacing *L* is 19*b*. But a rotation of $\omega=30°$ would imply a dislocation spacing of only *L=2b*. Clearly, the 30° rotation of the OR at the (111) Pb/Al interface exceeds the limit of the dislocation description and is therefore outside the range where strain is a reasonable indicator of the difference in interfacial energies. Against this background, it is remarkable that the systematic behavior seen in Table 1 can be understood in terms of transformation strains as the contribution best able to distinguish between interfacial energies in these systems.

*6.7 Summary of interface structures*

Within each column of Table 1, the orientation relationship *OR S* changes continuously so as to maintain the interface as an unrotated plane containing the eigenvector with the smaller strain. Since the directions of the strain are independent of the misfit, the interfaces maintain the same geometry as the mismatch changes within each row of the table. But the magnitude of the strain varies across the entire Table. Within each column, the strain changes with surface orientation, and across columns, the range of strains changes with mismatch. Our analysis indicates systematic behavior as long as the strain remains within roughly ±7%, but irregular behavior sets in when the strain goes outside this limit, such as for (322), (211) and (111) Al substrates in Pb/Al. For the former two substrate orientations, the film adopts the OR demanded by the 1D model (edge-to-edge matching), accommodating the difference in plane spacing by stacking faults. Interestingly, these faults are spaced periodically - a fault at every step on (322) substrates, forming the 12R structure, and at every other step on (211) substrates, forming the 9R structure. However, regardless of their specific arrangement, the role of stacking faults as a means of relieving the strain in the interface is a central feature, becoming more important for larger strains.

## 7. Conclusions

Using MD simulations, we have investigated the effect of mismatch on heteroepitaxial thin film growth in FCC/FCC systems. The three film/substrate systems studied covered a significant range of lattice mismatch, from 12.6% for Ag/Cu, to 16.2% for Ag/Ni, to 21.8% for Pb/Al. Six substrate orientations were selected for each system: (100), (511), (311), (211), (322) and (111), covering a range of 55°. All 18 interfaces were equilibrated by MD and relaxed by MS using Embedded Atom Method potentials. The results show that the ORs change gradually with substrate orientation from the oct-cube OR on (100) substrates to the cube-twin OR on most of the (111) substrates. This is in agreement with previous experimental results and simulations on Ag/Ni [2,4]. The simulations show that the ORs for all three systems are essentially identical, independent of lattice mismatch. The interfaces exhibit a step-and-terrace structure with a set of densely-packed planes meeting edge-to-edge at the steps. Most of the films display stacking faults emanating from some of the steps, and the density of these defects increases with mismatch. The results were interpreted in terms of two models based on 2D transformation strains and 1D edge-to-edge matching. While the ORs generally agree with the 2D model, all the interface structures are in accord with the 1D model. It is found that these differences in the predicted structures and orientations can be reconciled by elastic distortions and the presence of stacking faults at the interfaces.

Although this study is confined to three simple binary systems with limited solubility, there is no reason why similar behavior should not occur in more complex systems, including metals on oxide or semiconductor substrates.

**Acknowledgments**

PW wishes to acknowledge use of the computational resources of the National Energy Research Scientific Computing Center, which is supported by the Office of Science of the U.S. Department of Energy under Contract No. DE-AC02-05CH11231. DC wishes to thank the Agence Nationale de la Recherche for support of her research under grant ANR-GIBBS-15-CE30-0016. Work at the Molecular Foundry was supported by the Office of Science, Office of Basic Energy Sciences, of the U.S. Department of Energy under Contract No. DE-AC02-05CH11231.